\documentclass{aa}
\usepackage{txfonts}
\usepackage{psfig}
\usepackage{natbib}
\bibpunct{(}{)}{;}{a}{}{,}

\begin{document}

\def\apss{Astrophysics \& Space Science}
\def\eps{\varepsilon}
\def\aap{A\&A}
\def\apj{ApJ}
\def\apjl{ApJL}
\def\mnras{MNRAS}
\def\aj{AJ}
\def\nat{Nature}
\def\aaps{A\&A Supp.}
\def\e{{\rm e}}
\def\me{m_\e}
\def\lesssim{\mathrel{\hbox{\rlap{\hbox{\lower4pt\hbox{$\sim$}}}\hbox{$<$}}}}
\def\gtrsim{\mathrel{\hbox{\rlap{\hbox{\lower4pt\hbox{$\sim$}}}\hbox{$>$}}}}

\def\del#1{{}}

\title{Does circular polarisation reveal the rotation of quasar engines?}
\titlerunning{Does circular polarisation reveal the rotation of 
quasar engines?}
\author{Torsten A. En{\ss}lin}
\authorrunning{T. A. En{\ss}lin}
\institute{Max-Planck-Institut f\"{u}r Astrophysik,
Karl-Schwarzschild-Str.1, Postfach 1317, 85741 Garching, 
Germany} 
\date{Received 6. November 2002 / Accepted 6. February}

\abstract{Many radio sources like quasars, blazars, radio galaxies,
and micro-quasars exhibit circular polarisation (CP) with surprising
temporal persistent handedness. As a possible explanation we propose
that the CP is due to Faraday conversion (FC) of linear polarisation
(LP) synchrotron light which propagates along a line-of-sight (LOS)
through twisted magnetic fields. The rotational nature of accretion
flows onto black holes naturally generates the required magnetic twist
in the emission region, independent of whether it is a jet or an
advection dominated accretion flow (ADAF). The expected twist in both
types of flows is of the order of what is required for optimal CP
generation. This scenario requires that Faraday rotation (FR) is
insignificant in the emission region. Although this is an assumption,
it relaxes constraints on the plasma parameters, that were given in
scenarios which rely on FR, since there the strength of FR can not be
too far from the optimum.  The proposed mechanism works in electron-positron
($\e^\pm$) as well as electron-proton ($\e/{\rm p}$) plasma. In the
latter case, the emission region should consist of individual flux
tubes with independent polarities in order to suppress too strong FR.
The predominant CP is expected to mostly counter-rotate (rotation is
measured here in sky-projection) with respect to the central engine in
all cases (jet or ADAF, $\e^\pm$ or $\e/{\rm p}$ plasma).  If the
proposed mechanism is indeed operating, it will allow to measure the
sense of rotation of quasar engines.  The engine of SgrA$^*$ is then
expected to rotate clockwise and therefore counter-Galactic, as do the
young hot stars in its vicinity, which are thought to feed SgrA$^*$ by
their winds.  Similarly, we expect the microquasars SS~443 and
GRS~1915+105 to rotate clockwise. Generally, sources with Stokes-$V<0$
($V>0$) are expected to rotate clockwise (counter-clockwise) in this
scenario.
 } \maketitle

\section{Introduction\label{sec:intro}}

The long lasting interest in CP from quasar-like systems \citep[e.g.][
in the following J\&O'D]{1977ApJ...214..522J} increased recently due
to the detection of several new CP sources as SgrA$^*$
\citep{1999ApJ...523L..29B, 1999ApJ...526L..85S}, M81$^*$
\citep{2001ApJ...560L.123B}, the micro-quasars SS~433 and GRS~1915+105
\citep{2000ApJ...530L..29F, 2002MNRAS.336...39F}, low luminosity AGNs
\citep{2002ApJ...578L.103B}, and many additional powerful quasars
\citep{1998Natur.395..457W, 1999AJ....118.1942H} and blazars
\citep{2001ApJ...556..113H}.  In the following we will summarise these
systems under the general term quasar, assuming that a similar
mechanism produces CP in most of them. The level of CP is $\sim 1\%$
and below, usually (but not in the case of SgrA$^*$) much lower than
the level of LP.  On the one hand CP is highly time variable, but on
the other hand in many sources it exhibits a very persistent
rotational sense (per source and at nearly all frequencies where
detected), which is constant on timescales of decades
\citep{1984MNRAS.208..409K, 1999AJ....118.1942H, 2002ApJ...571..843B},
although exceptions exist. This is far in excess of the dynamical
timescale of the central quasar engine from which the emission
originates. 

In these sources synchrotron radiation from relativistic electrons
produces LP, but only a very small and usually negligible amount of
CP. A number of mechanisms have been proposed to explain the observed
CP, among which FC of LP to CP seems to be the most likely process
\citep{2002MNRAS.336...39F,2002A&A...396..615M}. \cite{2002A&A...388.1106B}
and \cite{2002ApJ...573..485R} (B\&F and R\&B in the following)
provide a good introductions into the matter, a detailed discussion of
the various CP generation mechanisms, and further references. For a
discussion of scintillation models for CP generation
\cite{2002PASA...19...43M} should be consulted.

FC of synchrotron emission is a two-step process, since the emitted
polarisation state has to be rotated before it can be Faraday
converted into circular polarisation. This can happen by FR, as
discussed by e.g. B\&F and R\&B and used as a starting point of this
work, or it can be done by a systematic geometrical rotation of the
magnetic field along the LOS, as originally proposed by
\cite{1982ApJ...263..595H} and discussed here.

The motivation for the standpoint adopted in this Letter, the
assumption that the rotation of LP is a geometric effect rather than
FR, is the observational fact that CP changes are very rare, much more
rare than LP rotator events, and that a predominant CP sign seems to
be persistent in many sources.  As explained in the following, this
observational fact requires in the FR based models a constant magnetic
polarity in the emission region over timescales far in excess of the
dynamical timescale of the central engine, since the sign of CP
depends on the magnetic field polarity in such models.  In the
geometrical model discussed here, the CP sign is fully determined by
the rotational sense of the central engine. Therefore the persistence
of the predominant CP sign would -- in this picture -- be a natural
consequence of angular momentum conservation.

\section{Faraday conversion\label{sec:conv}}
\subsection{Homogeneous magnetic fields\label{sec:homo}}
We start our discussion with the case of a homogeneous magnetic field,
and follow the notation of J\&O'D, B\&F, and R\&B, which should be
consulted for details and references.  The evolution of the Stokes
polarisation parameters $I,Q,U,V$ along a given LOS (here defining the
$z$-axis) is governed by
\begin{equation}
\label{eq:master}
\frac{d}{dz} \left(
\begin{array}{c}
I \\ Q\\ U\\ V
\end{array}
 \right) = 
\left( 
\begin{array}{c}
\eps_I \\ \eps_Q\\ 0 \\ \eps_V
\end{array}
\right)
- \left( 
\begin{array}{cccc}
\kappa_I & \kappa_Q  & 0         & \kappa_V \\
\kappa_Q & \kappa_I  & \kappa_{\rm F}  & 0 \\
0        & -\kappa_{\rm F} & \kappa_I  & \kappa_{\rm C} \\
\kappa_V & 0         & -\kappa_{\rm C} & \kappa_I 
\end{array}
\right)  \left( 
\begin{array}{c}
I \\ Q\\ U\\ V
\end{array}
\right)\!\!,
\end{equation}
where $\eps_{I/Q/U/V}$ is the emissivity and $\kappa_{I/Q/U/V}$ the
absorption coefficient of radiation with Stokes parameter $I/Q/U/V$. A
special coordinate frame is adopted, in which the sky-projected
magnetic field is parallel to the $y$-axis ($\vec{B} = (0, B_y,B_z)$),
which leads to vanishing synchrotron emission and absorption
coefficients for Stokes-$U$ \footnote{LP, $\pi/4$ inclined to the
sky-projected magnetic fields.}. The emission and absorption
coefficients for Stokes-$V$ \footnote{CP, clockwise or negatively
rotating in sky-projection for $V>0$, counter-clockwise or positively
rotating for $V<0$.} are much smaller than the ones for Stokes-$Q$
\footnote{LP, perpendicular ($Q>0$) or parallel ($Q<0$) electric
oscillation with respect to the sky-projected magnetic field.} -- if
they are not also zero as would be the case for an $\e^\pm$
plasma\footnote{\cite{1984ASS...100..227V} proposed a model in which
relativistic beaming allows to see the small amount of circularly
polarised synchrotron emission, whereas the linear polarisation is
suppressed by cancelling in a helical magnetic field configuration
seen face on. Note, that in this model the CP sign depends on the
magnetic polarity.}.
The FR coefficient reads
\begin{equation}
\kappa_{\rm F} = - \frac{\tilde{\rho}_\e\, e^2\, B_z}{\pi\, \me^2\, c^4}
\lambda^2, 
\end{equation}
where $\tilde{\rho}_{\e}$ gives the effective Faraday-active leptonic
charge density.  For a non-relativistic plasma $\tilde{\rho}_{\e}$ is
identical to the leptonic charge density $\rho_{\e} = e\,(n_{\e^+} -
n_{\e^-})$, and for a relativistic plasma $\tilde{\rho}_{\e} \propto
\rho_{\e}$ with a constant of proportionality smaller than
one.\footnote{$n_{\e^\pm}$ is the $\e^\pm$ number density
respectively. In the relativistic case an identical spectral
distribution of the two species is assumed, otherwise they enter
$\tilde{\rho}_\e$ with different proportionality factors.}  The sign
of FR depends on the sign of $\rho_\e$ and on the direction of the
magnetic field component along the LOS ($B_z$).
The FC coefficient
\begin{equation}
\kappa_{\rm C} = - \frac{\tilde{n}_{\e}\,e^4\,B_y^2}{4\,\pi^2\, \me^3\,c^6}
\lambda^3 
\end{equation}
is sensitive to the total number of leptonic charge carriers
$\tilde{n}_\e \propto n_\e = n_{\e^+} + n_{\e^-}$ (with a
proportionality constant of one in the non-relativistic and smaller
than one in the relativistic case) and the square of the perpendicular
field strength $B_y^2$. Since both numbers are always positive,
$\kappa_{\rm C}$ is always negative (or zero).
Since both Faraday coefficients strongly increase with wavelength
their importance is largest at lowest frequencies, possibly explaining
why CP is often detected close to the synchrotron-self-absorption
frequency. The relatively flat observed $V$ spectrum of SgrA$^*$
\citep[compared to the $I$ spectrum,][]{2002ApJ...571..843B} may
indicate that for too strongly self-absorbed sources CP production
becomes inefficient, since it is expected to happen only within the
$\tau = \int dz\,\kappa_{I} \le 1$ region.
In an $\e/{\rm p}$ plasma with diagonal magnetic fields ($|B_y| \sim
|B_z|$) the FR coefficient is usually larger in magnitude compared to
the FC coefficient, except for extremely high field strength or
extremely low frequencies. Both are not expected in our case.

Since there is no direct conversion of $Q$ synchrotron emission to $V$
(the corresponding matrix element in Eq. \ref{eq:master} is zero) CP
has to be produced via a two-step conversion in this setting: After FR
of $Q$ into $U$ it is further converted by FC to $V$. The sign of $V$
depends only on the sign of $\kappa_{\rm F}$ and therefore on the
polarity of the magnetic field with respect to the LOS. In a charge
symmetric $\e^\pm$ plasma ($\rho_{\e^\pm}=0$) FR vanishes, and this
mechanism is unable to operate. On the other hand the total Faraday
depth ($\tau_{\rm F} = \int dz\,\kappa_{\rm F}$) should not be too
large within the converting region, otherwise the continued FR changes
the sign of $U$, which leads to the production of $V$ with the
opposite sign and therefore cancelling of CP. An effective CP
production can be achieved if both $\tau_{\rm F}$ and $\tau_{\rm C} =
\int dz\,\kappa_{\rm C}$ are simultaneously of the order one, since $V
\approx \frac{1}{6}\, \tau_{\rm F} \,\tau_{\rm C}\, Q$ for $\tau_{\rm
F} \ll 1$ and $\tau_{\rm C} \ll 1$ within an optically thin
region\footnote{For an optically thick source, using only the region
up to $\tau =1$ is a reasonable approximation, leading to $V \approx
\frac{1}{6}\, \tau_{\rm F} \,\tau_{\rm C}\, Q/\tau^2$.  CP sign
changes can occur around $\tau\sim 1$
\citep{1977ApJ...214..522J,1977ApJ...215..236J}.  However, the
intensity of the CP flux with reversed sign is usually much below the
flux with the predominant sign. For that reason
\cite{1990MNRAS.242..158B} concluded that in inhomogeneous situations
such CP reversals are usually suppressed. Anyhow, here we concentrate on
the predominant sign of CP, and ignore possible complication which can
occur occasionally at special frequencies.}, and much less
than this if $\tau_{\rm F} \ge 1$ and/or $\tau_{\rm C} \ge 1$ due to
over-rotation and/or over-conversion.

In order to produce a significant $V$ component in the case of a
homogenous magnetic field either $\kappa_{\rm F}$ has to be fine tuned
with the help of a suitable field orientation\footnote{Ideally,
the field would be oriented mostly perpendicular to the LOS -- but
not perfectly, as B\&F point out by emphasising the advantage of a
helical field configuration.} or by a small but non-zero effective
leptonic charge density, since $\kappa_{\rm F} /\kappa_{\rm C} \propto
\varrho_\e/n_\e$. Both possibilities seem to be too contrived to allow
CP to be detected in a significant number of very different sources,
ranging from micro-quasars over low-luminosity AGNs to powerful
quasars.

\subsection{Stochastic magnetic fields}
B\&F and R\&B demonstrated that a significant CP production can be
achieved even in the case of a very large microscopic FR coefficient
if the magnetic field is mostly stochastic. The contributions of
$\kappa_{\rm F}$ with differing signs due to magnetic field reversals
can cancel each other, leaving only a small total Faraday depth
$\tau_{\rm F}$ due to an assumed weak mean field.  The strength and
sign of this mean field determines the strength and sign of the
resulting CP respectively. Although some fine-tuning is also necessary
in this scenario in order to simultaneously have the right order of FR
and FC, the larger number of free parameters (field strength and
coherence length, ratio of mean to stochastic field components,
$\tilde{\rho}_\e$, $\tilde{n}_\e$) provides a sufficiently large
parameter space to make this scenario a plausible explanation for the
observed CP for a wide range of objects. R\&B showed that this
approach can produce CP in the case of an $\e/{\rm p}$ and also in the
case of a (charge-asymmetric) $\e^\pm$ plasma.

However, in order to explain the long-term stability of the observed
sign of $V$ it has to be assumed that the weak magnetic mean flux has
to retain its polarity. This might surprise, since the dynamical time
scales in several of the observed systems are orders of magnitudes
smaller than the time-scale over which stability of the sign of $V$
could be established\footnote{E.g. SgrA$^*$ and GRS~1915+105.
SgrA$^*$ exhibited a stable $V$ sign on a timescale (20 years) which
is 5 orders of magnitude longer than the dynamical timescale of the
accretion flow close to the black hole ($\sim$ h).  GRS~1915+105 was
observed to exhibit a LP-rotator event, but no change in the sign of
CP \citep[][ who described an indication of a CP sign reversal to be
insignificant]{2002MNRAS.336...39F}. However, counter-examples seem to
exist, e.g. a CP sign reversal seems to be observed during the onset
of a radiation outburst of GRO J1655-40 in 1994
\citep{2002A&A...396..615M}.}.  Furthermore, the large fluctuations in
$V$ indicate strong temporal variations in the strength of the mean
field. Since the latter should be dynamically unimportant, its stable
polarity may be best explained if it results from an environmental
large scale field, which gets dragged into the central engine. An
alternative possibility was suggested by the anonymous referee: the
mean field may be a remnant field of a stochastic superposition of
many independent magnetic flux patches. The dynamical timescale for
the evolution of the mean field could be on timescales much larger
than the timescale of the evolution of an individual magnetic patch.

\begin{figure}[t]
\begin{center}
\psfig{figure=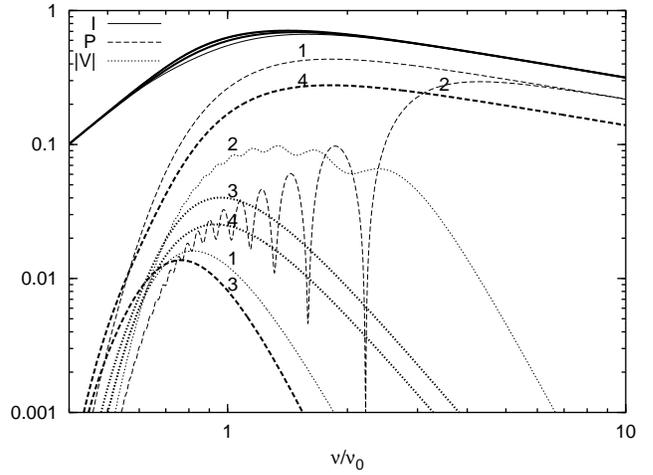,width=0.5 \textwidth,angle=0} %
\end{center}
\caption[]{\label{fig:example} Model spectra from numerical solutions
of the radiative transfer along a LOS with homogeneous
parameters. Shown is the total intensity $I$, the linearly polarised
flux $P = \sqrt{Q^2 + U^2}$, and the circular polarisation $|V|$. The
sign of $V$ is always negative in the given examples, in accordance
with the argumentation in the text.  The LOS is characterised by
$\eps_I = (\nu /\nu_0)^{-\alpha}$, $\eps_Q = (\alpha+1)/(\alpha + 5/3)
\,\eps_I$, $\eps_U = \eps_V = 0$, $\kappa_I = (\nu
/\nu_0)^{-\alpha -5/2}$, $\kappa_Q = (\alpha+1)/(\alpha + 5/3)\,
\kappa_I$, $\kappa_{\rm R} = \tilde{\kappa}_{\rm F}\,(\nu /\nu_0)^{-2}
+ \Omega$, and $\kappa_{\rm C} = \tilde{\kappa}_{\rm C}\,(\nu
/\nu_0)^{-3}$. The depth of the source is assumed to be $\Delta z =1$,
so that $\tau =1$ for $\nu = \nu_0$.  The thin lines (model 1 \& 2)
are FR based models, whereas the thick lines (model 3 \& 4) are
magnetic twist based models. The model parameters are $\alpha = 0.5$
in all cases, $\tilde{\kappa}_{\rm F} = 1, 30, 0, 0$, $\Omega =
0,0,2\pi, \pi$, and $\tilde{\kappa}_{\rm C} = -0.3,-10,-1,-0.3$
respectively for model 1,2,3,4. Note, that the parameters were chosen
to illustrate the variety of possible spectra, to lead to an effective
CP production, and to provide a readable figure. Therefore the
parameters do not necessarily represent typical situations, and in
some cases they are even counter examples to trends discussed in the
text (e.g. model 2 exhibits strong CP due to an optimally chosen $\kappa_{\rm
C}$).}
\end{figure}

\subsection{Twisted magnetic fields}

The formalism of Sect. \ref{sec:homo} can also be applied to the case
of inhomogeneous magnetic fields if all coefficients in
Eq. \ref{eq:master} become spatially dependent, and the FR coefficient
is replaced by
\begin{equation}
\kappa_{\rm F} \rightarrow \kappa_{\rm R} = \kappa_{\rm F} + \Omega.
\;\;\;\;\;\;\;\;\;\mbox{Here,}\; \Omega = -2\,d\phi_B/dz
\end{equation}
describes the rate of linear polarisation rotation along the LOS due
to the rotation of the coordinate frame, which was tied to the
sky-projected magnetic fields for convenience. $\phi_B$ is the
position angle of the sky-projected magnetic field in a fixed
(non-varying) coordinate frame. For a purely stochastic magnetic
field, $\Omega$ is a random variable with zero mean. However, if there
is a systematic twist in the magnetic fields along the LOS $\Omega$
has a preferred sign. The sign of $\Omega$ does not depend on the
polarity of the field, only on the handedness of the twist along the
LOS. The effect of magnetic twist for CP production is mathematically
identical to that of FR, but only if a single frequency is regarded,
since FR is frequency dependent, whereas magnetic twist is not.
However, CP production is possible in a completely non-Faraday
rotating medium, e.g. as a charge symmetric $\e^\pm$ plasma. This is
illustrated in Fig. \ref{fig:example}, where simple model spectra for
FR based and magnetic twist based CP generation scenarios are
presented.

The requirement of $\tau_{\rm R} = \int dz\,\kappa_R\sim 1$ for
effective conversion translates into the requirement that the total
magnetic twist along a LOS within the converting region is of the
order of one (optimally it would be below $\pi/2$ in order to avoid
cancellation of $V$ contributions with opposite signs). It is argued
below that this condition can be fulfilled naturally. In order to have
magnetic twist being the dominant cause of $Q$ to $U$ conversion, FR
has to be sufficiently suppressed. We assume in the following that
$\tau_{\rm F} \ll 1$ holds within the optical thin part of the
emission region, and show that this naturally leads to the observed
sign persistence of $V$.  The suppression of FR could be due to a
vanishing leptonic charge density ($\rho_\e$) as expected for a
$\e^\pm$ plasma, or due to the presence of aligned, separate or nested
flux tubes (with a global large scale twist) with independently
oriented (but aligned) internal magnetic fields. As B\&F and R\&B
showed for the stochastic field case, this can strongly reduce the
Faraday depth also in the case of a large microscopic $\kappa_{\rm
F}$. In their cases, some fine-tuning was required since FR should be
somehow suppressed, but not completely, leaving to sufficient FR from
$Q$ to $U$. This fine-tuning is not required here, since the
geometrical rotation provides $Q$ to $U$ rotation even in the case of
fully suppressed FR.  $\kappa_{\rm C}$ is not affected by LOS
reversals of the magnetic field direction, since it only depends on
the invariant $B_y^2$.

\section{Circular polarisation and rotating flows\label{sec:cparf}}
\subsection{A jet scenario\label{sec:jet}}

\begin{figure}[t]
\begin{center}
\psfig{figure=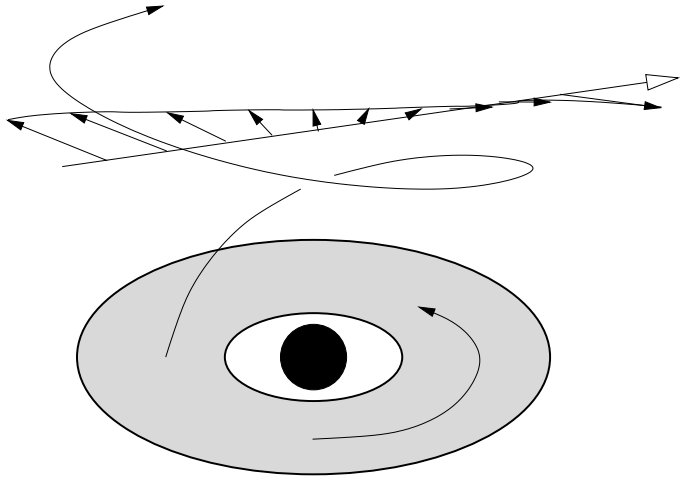,width=0.5 \textwidth,angle=0}%
\end{center}
\caption[]{\label{fig:jet} Sketch of a possible geometry of a jet
source. Shown is a positively rotating accretion disk around a central
black hole from which a jet is launched (not shown). A single helical
field line of the jet is drawn, and the magnetic field orientation
along a typical LOS is indicated, demonstrating the presence of
magnetic twist around the LOS.}
\end{figure}

Twisted magnetic fields are expected to be a natural ingredient of any
quasar central engine due to the rotational nature of the accretion
flow.  A helically twisted magnetic field is expected to be present in
jet outflows, especially if jets are magneto-hydrodynamically launched
\citep{1982MNRAS.199..883B}, but also otherwise, due to decreasing
rotation speed of the sideways expanding outflow.  The strength of the
twisting depends on the ratio of the rotation to the outflow velocity
(modulo beaming). In typical situations one would expect both
velocities to be of the same order of magnitude, leading to
$|\tau_{\rm R}| \sim 1$, as required for an efficient conversion in
the absence of FR. Further, $|\tau_{\rm R}| \le 2\,\pi$, where
equality corresponds to a full reversal of the field across the jet in
case of a completely wound up toroidal field. If the fields are very
toroidal ($\tau_{\rm R}\ge \pi$) an optical depth $\tau >1$ can help
to increase CP by restricting the effective FC volume to a fraction of
the LOS through the jet, and therefore suppressing CP cancellation
effects caused by a production of $V$ of both signs.

A typical geometry is sketched in Fig. \ref{fig:jet} for an
approaching jet, which is rotating positively (in sky-projection). The
rotation sense of the sky-projected magnetic fields seen by a photon
flying from its emission point (where the synchrotron process provided
it with $Q>0$) to the observer is therefore also positive, leading to
$\Omega = -2d\phi_B/dz <0$ in this geometry. Thus, in our limit
$\kappa_{\rm F} \ll \Omega$ we obtain $V\propto \Omega\,\kappa_{\rm
C}\,Q\ge 0$ since $\kappa_{\rm C} \le 0$ always.  In this picture a
positively rotating approaching jet emits negatively (clockwise)
rotating circularly polarised light.  Since the counter jets should
exhibit an oppositely twisted magnetic field, it should produce CP of
the opposite sign.  However, the approaching jet dominates the total
emission due to relativistic beaming so that its CP emission dominates
the CP of both jets. We can conclude that the rotation of the received
photons of a synchrotron emitting jet source are expected to be
retrograde to -- and therefore reveal -- the sense of rotation of the
central engine, which is the rotation of the accretion disk and/or the
black hole.

\subsection{An ADAF scenario\label{sec:adaf}}

\begin{figure}[t]
\begin{center}
\psfig{figure=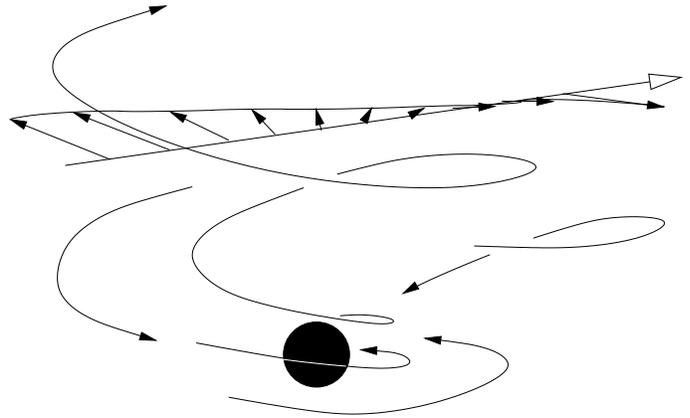,width=0.5 \textwidth,angle=0}%
\end{center}
\caption[]{\label{fig:adaf} Sketch of a possible ADAF onto a black
hole. The flow direction and a single magnetic field line are
shown. The converging and spiralling flow should imprint a helical
twist into the magnetic field, which has exactly the same handedness as
the one expected from a similarly rotating outflow, as shown in
Fig. \ref{fig:jet}. The twist of the magnetic field along a LOS (also
shown) has therefore the same sign in both models.}
\end{figure}

In an ADAF \citep{1994ApJ...428L..13N} CP can also result from
converted synchrotron emission, if the ADAF region produces
synchrotron emission. The handedness of the magnetic twist is expected
to be the same as the one in jets (if FR is suppressed in ADAFs) for the
following reason: the ADAF is a converging flow, that is spinning up
due to angular momentum conservation, whereas the jet is an expanding
down-spinning flow. The way magnetic fields should be dragged in both
cases has therefore the same handedness with respect to the spin axis
of the system (see Fig. \ref{fig:adaf}). This leads to the same
handedness of the magnetic twist in the jet and ADAF case (on the
upper and lower hemisphere of the black hole respectively).  If the
received flux is dominated by the upper (nearer) side of the ADAF, due
to optical depth effects (CP sources are often self-absorbed), or due
to the black hole absorption of photons from the lower ADAF
hemisphere, a net CP results, which counter-rotates to the ADAF. The
fractional CP of an ADAF may be comparable to that of a jet since
$|\tau_{\rm R}| \sim 1$ and $|\tau_{\rm R}|\le 2\pi$ should hold for
the same reasons as in the jet case.  CP is therefore not necessarily
uniquely a jet signature.

\section{Discussion and outlook}

Twisted magnetic flux can produce circularly polarised synchrotron
emission due to FC. If FR is sufficiently suppressed the expected
sense of rotation of the emission is expected to be opposite to the
one of the central engine, independent of the question whether the
emission results form a synchrotron jet, or from a synchrotron
emitting ADAF, and independent of the polarity of the magnetic fields.
The FR suppression can be -- as discussed by B\&F and R\&B -- due to a
charge symmetric $\e^\pm$ plasma, in which FR is absent, or due to
small-scale magnetic field polarity changes which produce cancelling
FR contributions, but do not affect FC if the different flux tubes
show alignment.  Such polarity changes could be present if the
magnetic field is organised in many helically aligned flux tubes of
independent magnetic field polarity which originated in differently
magnetised patches of the accretion disk.

This model is not in conflict with other proposed generation
mechanisms for CP (as e.g.  stochastic Faraday rotation with a small
mean magnetic field as proposed by B\&F and R\&B), rather, it
supplements them. Several of its strengths listed below are also
present in other models. However, a few of the strengths (No. 1., 2.,
and 6.) seem to be slight advantages, potentially suggesting that it
could be the dominant mechanism in several cases:
\begin{enumerate}
\item The sign of the CP does not depend on the presence of a mean
field, but on geometrical properties of the flow pattern in the
central engine of a quasar, which are fixed by the system's angular
momentum. This may explain naturally the observed sign persistence of
the CP over periods which exceed the dynamical timescales of the
system by orders of magnitude.
\item The requirement that the rotation of the angle between LP and
projected magnetic field is of the order one (not much higher, since
over-rotation reduces CP, not much lower in order to have FC
operating) within the optically thin part of the emission region, is
naturally fulfilled by the geometrical properties of jet and ADAF
flows. This requirement implies some level of fine-tuning in other
models where the rotation is due to FR.
\item CP is still converted LP in this scenario, so that the
variability of CP should exceed the one of LP as observed.
\item The mechanism can work in jets and in ADAFs.
\item It is able to produce CP in an $\e^\pm$
and in an e/p plasma.
\item The relatively large robustness of the mechanism to variations
in the underlying source properties -- as composition of the Faraday
converting plasma and nature of the emission region (jet or ADAF) --
may explain why CP appears in a large variety of very different
systems, such as micro-quasars, low luminosity AGNs, and powerful quasars.
\item In cases where CP exceeds LP (e.g. SgrA$^*$), an outer Faraday
screen, e.g. located in a mixing layer around an $\e^\pm$ jet, could
provide the depolarisation \citep[e.g.][]{2001ApJ...556..113H}.
Alternatively, a $\pi/2$ rotation of the sky-projected magnetic field
throughout the visible emission region (the full optical thin jet, or
the $\tau \le 1$ region) could remove any observable LP, but leave a
measurable amount of CP.
\end{enumerate}
There are conditions, under which the here proposed relation between
the (predominant) rotational sense of CP and the rotation of the
central engine is not valid anymore. Whenever the average FR is
stronger than the twist in the magnetic fields along the LOS, the sign
of the average magnetic field will determine the sign of CP as in the
models of B\&F and R\&B. This may occur temporarily, if thermal plasma
gets mixed into an $\e^\pm$ jet. If the emission region contains an
e/p plasma, the occurrence of a mean magnetic field -- either as a
statistical fluctuation, or due to some physical reason -- can
increase the average Faraday depth and lead to a CP reversal. External
CP generation from LP due to environmental scintillation also can mask
the intrinsic CP signature of a quasar. Also within the scenario in
which CP is due to bending in magnetic fields, CP sign reversals are
possible, whenever the magnetic twist within the emission region is
not linked to the central engine rotation. This may happen temporarily
in violent shocks in a jet outflow, and would suggest that CP sign
reversals of this origin are likely accompanying emission outbursts of
the source\footnote{This may explain the observed CP sign reversal in
GRO J1655-40 in during an outburst in 1994, as reported by
\cite{2002A&A...396..615M}}.  The CP should also change if the
rotational sense of the central engine is changed, e.g. due to a black
hole merging event, or due to a massive infall of fresh material onto
the accretion disc.  Note, that the rarity of CP sign reversals
indicate that the above discussed conditions should be exceptional, if
CP is due to the here proposed mechanism.

The proposed mechanism provides testable predictions: 
\begin{enumerate}
\item The sign of CP from a quasar measures the rotation direction
of the quasar's engine. The electric vector of radio emission should
rotate retrograde in the sky-plane with respect to the engine
rotation. Positive $V$ polarisations implies therefore positive $=$
counter-clockwise rotation (on the sky) of the engine. 
\item Thus, we expect SgrA$^*$, which exhibited $V<0$ during the last
20 years, to rotate clockwise. This is retrograde with respect to the
galactic rotation and the rotation of the molecular gas cloud in the
galactic centre. However, it has the same rotation sense as the
population of young hot HeI stars in its direct vicinity
\citep{2000MNRAS.317..348G}, which were proposed to feed SgrA$^*$ via
their stellar winds \citep[][ for discussion and
references]{2001dscm.conf..291G}. Conservation of angular momentum in
this wind should lead to a retrograde (clockwise) rotating accretion
disk. Alternatively, the spin of the central black hole may determine
the twist of the magnetic fields in the emission region.
\item We expect the microquasars GRS~1915+105 and SS~433 (both exhibit
$V<0$) also to rotate clockwise, and we could give corresponding
expectations for the other sources with detected CP.
\item The predominant sign of CP should be temporally constant. If the
model is correct, quasar-like CP sources should only exhibit CP flips
under exceptional circumstances, as discussed above.
\item CP from the counter jet should have opposite handedness than CP from
the approaching jet. This prediction may allow a discrimination of the
presented model from FR based scenarios, since in the latter the same
predominant CP sign is expected from both jets for a dipole-like mean
field component. 
\item There are possibly also spectral signatures, which allow to
discriminate the different models. E.g. in FR based scenarios
frequency dependent LP oscillations can occur, as illustrated by model
2 in Fig. \ref{fig:example}, whereas this is impossible in the
magnetic twist based scenarios. However, due to cancellation effects
in the superposition of many slightly different LOSs such oscillations
likely appear as a strong suppression of LP. Due to the high
complexity of the resulting polarisation spectra general rules are not
obvious and would require further investigations. In that context it
is interesting to note that Beckert (in preparation) recently
presented new model calculations for SgrA$^*$. In order to
simultaneously explain new high frequency LP measurements of SgrA$^*$
model parameters seem to be favoured for which magnetic twist is more
important for CP production than FR.
\end{enumerate}
If the scenario of CP production proposed here could be demonstrated
to be operatively in quasars, it would give us a tool to measure the
sense of rotation of the most powerful and enigmatic engines of the
universe -- the violent matter flows in the direct vicinity of black
holes -- by simply looking at the spin of the received photons.

\begin{acknowledgements}
I acknowledge useful discussions with F.~Meyer, S.~Heinz, H.~Spruit,
T.~Beckert, H.~Falcke, and C.~Pfrommer. The manuscript benefited from
several critical comments of an anonymous referee. After submission of
this paper I became aware that on a recent conference ({\it The
Central 100 Parsecs}, Hawaii, Nov. 3-8, 2002) T.~Beckert presented new
model calculations of the polarisation spectra of SgrA$^*$ in which a
helical field geometry plays the the same role as in this work.
\end{acknowledgements}



\end{document}